\documentclass[reprint,amsmath,amssymb,aps,]{revtex4-1}

\usepackage{graphicx}
\usepackage{dcolumn}
\usepackage{bm}
\usepackage{subfigure}
\usepackage{color}
\begin{document}

\preprint{APS/123-QED}

\title{Link Prediction in evolving networks based on the popularity of nodes}

\author{Tong Wang$^1$}

\author{Ming-yang Zhou$^{1\ 2}$}
\email{zhoumy2010@mail.ustc.edu.cn}

\author{Zhong-qian Fu$^1$}%

\affiliation{$^{1}$Department of Electronic Science and Technology, University of Science and Technology of
China, Hefei 230027, P. R. China.\\
 $^{2}$Guangdong Province Key Laboratory of Popular High Performance Computers, College of Computer Science and
 Software Engineering, Shenzhen University, Shenzhen 518060, P. R. China}

\date{\today}

\begin{abstract}
Link prediction aims to uncover the underlying relationship behind networks, which could be utilized to predict the missing edges or identify the spurious edges, and attracts much attention from various fields. The key issue of link prediction is to estimate the likelihood of two nodes in networks. Most current approaches of link prediction base on static structural analysis and ignore the temporal aspects of evolving networks.
Unlike previous work, in this paper, we propose a popularity based structural perturbation method (PBSPM) that characterizes the similarity of an edge not only from existing connections of networks, but also from the popularity of its two endpoints, since popular nodes have much more probability to form links between themselves. By taking popularity of nodes into account, PBSPM could suppress nodes that have high importance, but gradually become inactive. Therefore the proposed method is inclined to predict potential edges between active nodes, rather than edges between inactive nodes. Experimental results on four real networks show that the proposed method outperforms the state-of-the-art methods both in accuracy and robustness in evolving networks.

\end{abstract}


\maketitle

\section{Introduction}

Many complex systems in society and nature can be modeled as complex networks, with individuals denoted as nodes
and relations as links, respectively \cite{albert2002statistical,dorogovtsev2002pseudofractal}. A significant concern about complex networks is link prediction that aims at estimating the likelihood of two unconnected nodes based on obtained information \cite{liben2007link,wang2015link}. In protein-protein interaction experiments in cells, only strong relations between proteins could be detected by limited precision of equipments. Measuring every interaction between all pair proteins is prohibitive due to quadratical increasing experimental cost as the size of proteins \cite{mamitsuka2012mining,cannistraci2013link}. An appropriate approach is to evaluate the likelihood of potential relations and specifically test non-existing relations with high likelihood. In social networks, if two persons have many common friends, they would build friendship in the near future with high probability, which could be utilized to uncover lost friends or predict future friends \cite{kossinets2006effects,schifanella2010folks,Benchettara2010A}. Besides further extensive applications also includes personalized recommendation in e-commerce \cite{Zhou2010Solving,He2015Predicting}, aircraft route planning study and traffic dynamics in Internet topology \cite{yan2006efficient,watts1998collective}, and so on. Therefore link prediction attracts numerous researchers in various fields covering from biology to sociology and others \cite{clauset2008hierarchical,Yin2010LINKREC,guimera2009missing,zhang2013correction}.

The key issue of link prediction is evaluating the likelihood of potential edges, based on which we could rank the potential edges and edges in the top of ranking list are predicted as underlying or future edges. To achieve this, traditional
attribute based methods measure the existent likelihood of links by learning how many common features (e.g. common hobbies, ages, tastes, geographical locations) the two
nodes share \cite{lin1998information}. However this kind of methods suffer from the inaccessible and unreliable information of nodes due to privacy policy in real scienario \cite{lu2011link}.
Luckily the development of complex network provides a new path to solve the problem, in which only network topological structure is required regardless of privacy information. When evaluating the similarity between nodes, according to the structure differences, structure based methods could be classified into three categories: local methods, global methods and Quasi-global methods.
Local similarity is mainly based on common neighbors, such as the most well-known Common Neighbor (CN) index that counts the number of common neighbor nodes \cite{newman2001clustering}, Adamic-Adar (AA) index and Resource Allocation (RA) index that depress the large-degree neighbor nodes \cite{adamic2003friends,zhou2009predicting}. Global similarity emphasizes the global topology information of network, such as the Katz index counting all of the paths between two nodes \cite{katz1953new}. Quasi-global similarity is a well trade-off of local similarity methods and global similarity methods, such as the Local Path (LP) index that only considers the short paths in Katz index \cite{zhou2009predicting}, Local Random Walk (LRW) index that focuses on the limited random walk in local area \cite{liu2010link}. Beyond that, some algorithms based on maximum likelihood and other exquisite models have been proposed. Clauset \emph{et al.} proposed Hierarchical Structure Model which presents well performance in hierarchical networks by using a dendrogram \cite{clauset2008hierarchical}. Guimer\`{a} \emph{et al.} developed a Stochastic Block Model to identify the missing and spurious links, and obtained reconstructions of observed networks \cite{guimera2009missing}. Liu \emph{et al.} proposed a Fast Probability Block Model in which links are created preferentially to promote the clustering coefficient of communities, with the advantages of low computation complexity and high accuracy \cite{liu2013correlations}. L\"{u} \emph{et al.} proposed the Structural Perturbation Method based on the structural consistency that features of networks will not change sharply before and after a random addition or removal of links, presenting high accuracy and robustness in real-world networks \cite{lu2015toward}.

Unlike previous work that predict potential links mostly based on static networks, we propose a hypothesis that evolving of future links is not only determined by observed network topology, but also influenced by popularity of nodes. Important nodes in observed networks would attract more fresh edges according to preferential attachment, but some of them may become unpopular and fresh edges would avoid these nodes. That is to say, the ability of nodes to attract fresh nodes is decided by both its current importance and popularity. Then we propose a popularity based structural perturbation method to predict future edges. Comparing with other traditional approaches, experimental results on four real-world networks show that popularity based structural perturbation method (PBSPM) outperforms the other methods in accuracy with enhancement at most 112.09\%.

\section{Popularity metrics}
Consider an undirected and unweighted network $G(V,E)$, where $V$ and $E$ represent the set of nodes and links, respectively. Multi-links and self-loops are not allowed.
In real scenario, the networks always evolves toward a certain direction under the influence of specific events and other external factors.
For two nodes with same degree, one may connect its neighbors at early stage and not form connections later. While the other one develops most its connections at late stage.
Evidently the later node would attract more fresh edges with high probability in the near future. Inspired by this,
a straightforward approach to evaluate popularity of a node is counting the edges that it recently attracts.

Given a network with each edge having time-stamp that represents its entering time, we denote $k_i(t)$ the degree of nodes $i$ at time $t$.
In the next time span $T$, node $i$ would attract $\Delta k_i(t,T)$ new edges,
\begin{equation}\label{actualincrease}
  \Delta k_i\left( {t,T} \right) = {k_i}\left( {t + T} \right) - {k_i}\left( t \right).
\end{equation}

Note that degree increment in Eq. \ref{actualincrease} is determined by both $t$ and $T$. $\Delta k_i( {t,T})$ cannot reflect the relative activeness of node $i$, since even large degree nodes become inactive, they still attract more fresh edges than that of small degree ones due to preferential attachment. To solve the issue, for a dataset that spans time $t_{a}\sim t_{c}$, we divide it into $p_{older}$ fraction of old edges and $p_{fresher}$ fraction of fresh edges according to time sequence with time boundary $t_b$, $p_{older}+p_{fresher}=1$. The popularity of node $i$ is
\begin{equation}\label{activeness}
  {s_i} = \frac{\Delta k_i(t_b,t_c-t_b)}{\Delta k_i(t_a,t_c-t_a)}=\frac{{{k_{i,fresher}}}}{{{k_{i,all}}}},
\end{equation}
where $k_{i,all}$ and $k_{i,fresher}$ indicate the whole degree and fresher degree of node $i$. Note that $k_{i,fresher}\leq k_{i,all}$, thus $s_i\leq1$. If all degree of node $i$ locate in fresh set, ${s_i}=1$ means high popularity of node $i$. For another case that all degree of node $i$ locate in old set, node $i$ becomes inactive, ${s_i}=0$.
Therefore $s_i\in [0,1]$ and higher $s_i$ means higher popularity.

\section{Popularity based structural perturbation method}
In the section, we propose a hypothesis that the observed networks are determined by some latent attractors (e.g. similar hobbies, ages, sex, location). Given a network $G$ with adjacent matrix $A=(a_{ij})_{n\times n}$, for an attractor $x_k=[x_{k,1},x_{k,2},...,x_{k,n}]^T$ where $x_{k,i}$ represents the attractiveness of node $i$ for attractor $x_k$, inspired by configuration model, the probability $p_{ij}$ that an edge exists between two node $i$ and $j$ is proportional to $x_{k,i}\ast x_{k,j}$. Supposing that there are $m$ kinds of attractors, probability $p_{ij}$ equals to weighted influence of each attractor,
\begin{equation}\label{weightedattractor}
p_{ij}=\sum_{k=1}^{m} w_k\cdot x_{k,i}\ast x_{k,j},
\end{equation}
where $w_i$ is a tunable parameter to balance the relative influence of every attractor $x_k$. The problem is how to seek the optimal $w_i$ and $x_{k,i}$ that guarantee $p_{ij}$ approximating $a_{ij}$ at most. A special case is that $p_{ij}=1$ if $a_{ij}=1$, otherwise $p_{ij}=0$, where $a_{ij}$ is the element of adjacent matrix $A$. For optimal $w_i$ and $x_k$,
\begin{equation}\label{decompose}
A_p=(p_{ij})_{n\times n}=\sum_{k=1}^{m} w_k\cdot x_{k}\ast x_{k}^T.
\end{equation}

If $m=n$ in Eq. \ref{decompose} where $n$ is the size of the network, then Equation \ref{decompose} could be comprehended as the matrix decomposition, with $w_k$ and $x_k$ representing eigenvalues and eigenvectors respectively. In practice, many random connections exist in networks, L$\ddot{u}$ et al. proposed structural perturbation method (SPM) to reduce the influence of randomness. In SPM, a small fraction $p^H$ of edges $\Delta A$ is removed from the networks, $A=A^R+\Delta A$, adjacent matrix $A^R$ of the remaining networks is decomposed into
\begin{equation}\label{spectraldecomposition}
  {A^R} = \mathop \sum \limits_{k = 1}^N {\lambda _k}{x_k}x_k^T,
\end{equation}
where $\lambda_k$ and $x_k$ are the eigenvalues and eigenvectors of $A^R$, $|x_k|=1$. We could use $A^R$ to evaluate $A$ with
\begin{equation}\label{perturbatedmatrix}
  \tilde A = \mathop \sum \limits_{k = 1}^N \left( {{\lambda _k} + \Delta {\lambda _k}} \right){x_k}x_k^T,
\end{equation}
where $\Delta {\lambda _k}\approx \frac{{x_k^T\Delta A{x_k}}}{{x_k^T{x_k}}}$ is the coupling influence of $x_k$ on $\lambda _k$. $\tilde A$ actually is a special case of $A_p$ in Eq. \ref{decompose}. $w_k$ and elements of eigenvector $x_k$ represent weight difference and the attractiveness for attractor $x_k$ separately.

 As is mentioned above, the ability for node $i$ to attract new edges is determined by both latent attractors and popularity. To meet practice better, an advanced attractiveness $x_{k,i}'$ is proposed that
 \begin{equation}\label{futureattractiveness}
  {x}'_{k,i} = {x}_{k,i}\left( 1 + {\alpha *}{s}_{i} \right),
\end{equation}
where $\alpha$ is a tunable parameter. We later utilize $x_k'$ to substitute $x_k$ in  Eq. \ref{perturbatedmatrix} to predict future edges,
\begin{equation}\label{perturbatedmatrix2}
  \tilde A = \mathop \sum \limits_{k = 1}^N \left( {{\lambda_k} + \Delta {\lambda _k}} \right){x'_k}x_k'^T.
\end{equation}

Since Eq. \ref{spectraldecomposition} degenerates into Eq. \ref{decompose} if the size $m$ of attractors is less than $n$. According to theoretical analysis, suppose that $|\lambda_1|>|\lambda_2|>...>|\lambda_n|$ in Eq. \ref{spectraldecomposition}, we substitute $w_k$ and $x_k$ in Eq. \ref{decompose} with $\lambda_k$ and $x_k$ in Eq. \ref{spectraldecomposition}, similar to the same transition from Eq. \ref{spectraldecomposition} to Eq. \ref{perturbatedmatrix2}, we obtain
\begin{equation}\label{decompose2}
A'=(p_{ij})_{n\times n}=\sum_{k=1}^{m} ({{\lambda_k} + \Delta {\lambda _k}})\cdot x_{k}'\ast x_{k}'^T.
\end{equation}

Equation \ref{decompose2} reduces into Eq. \ref{perturbatedmatrix2} if $m=n$. In the following experiment, we firstly measure the performance of Eq. \ref{perturbatedmatrix2}, then show that we could reduce the calculation complexity by using only a few eigenvalues and eigenvectors, that is $m\ll n$ in Eq. \ref{decompose2}.

\section{EXPERIMENTAL METHODS}
\subsection{Experimental process}
For a dataset, we firstly divide it into 90\% training set $A^T$ and 10\% probe set $A^T$ based on timestamps attaching to edges, with training set and probe set containing old edges and fresh edges separately. Popularity of nodes are obtained by only training set. Training set is further divided into $p_{older}$ fraction of old edges and $p_{fresher}$ fraction of fresh edges according to time sequence, $p_{older}+p_{fresher}=1$. Then we utilize Eq. \ref{activeness} to calculate popularity.

Later we perturb training set by randomly removing a small fraction $p^H$ of edges $\Delta A$, $A^T=A^R+\Delta A$. Then $\tilde{A}$  and $A'$ could be obtained, in which $a'_{ij}$ represents the existent likelihood of link between node $i$ and $j$. Nonexisting edges with high score $\tilde{A}_{ij}$ and $a'_{ij}$ are chosen as potential future edges. All the experiments are the result of 10 independent simulations.

\subsection{Metric}
In this experiment, we choose Precision index as the metric to evaluate the accuracy of link prediction methods. Precision is defined as the ratio of links predicted accurately to all links selected\cite{Herlocker2004Evaluating}. In other works, if we select $Top-L$ links in the all ranked non-observed links and only ${L_r}$ links are predicted rightly in probe set $E^P$, the accuracy of predictor follows
\begin{equation}\label{Precision}
  Precision = \frac{{{L_r}}}{L}.
\end{equation}
Obviously, precision is sensitive to $L$ which is fixed at $\left| {{E^P}} \right|$ in this paper.

\subsection{Baselines}
For comparison, we briefly introduce five traditional algorithms based on all three kinds of structural similarity.
\par (1) Common Neighbors (CN) is the basic method which assumes that two endpoints tend to connect with each other if they have much more common neighbors.
\begin{equation}\label{CN}
  s_{xy}^{CN} = \left| {\Gamma \left( x \right) \cap \Gamma \left( y \right)} \right|,
\end{equation}
$\Gamma \left( x \right)$ is the set of neighbors of node $x$ and $\left| {\Gamma \left( x \right) \cap \Gamma \left( y \right)} \right|$ represents the set of common neighbors of $x$ and $y$.
\par (2) Adamin-Adar (AA) assumes that the contributions of common nodes are measured by the logarithm of reciprocal of their degrees.
\begin{equation}\label{AA}
  {s_{xy}^{AA}} = \mathop \sum \limits_{z \in \Gamma \left( x \right) \cap \Gamma \left( y \right)} \frac{1}{{\log {k_z}}},
\end{equation}
where $k_z$ denotes the degree of node $z$.
\par (3) Resource Allocation (RA) is similar with AA, regarding the reciprocal of common neighbors as the ability of transmission.
\begin{equation}\label{RA}
  {s_{xy}^{RA}} = \mathop \sum \limits_{z \in \Gamma \left( x \right) \cap \Gamma \left( y \right)} \frac{1}{{{k_z}}}.
\end{equation}
\par (4) Katz index, based on global information, counts all the paths connecting two endpoints with weakening the contributions of longer paths exponentially:
\begin{equation}\label{Kata}
  {s_{xy}^{Katz}} = \mathop \sum \limits_{l = 1}^\infty  {\alpha ^l} \cdot \left| {paths_{x,\;y}^{\left\langle l \right\rangle }} \right|.
\end{equation}
When $\left| \alpha  \right| < 1/{\lambda _{max}}$, it can be rewritten as:
\begin{equation}\label{Katz2}
  S = {\left( {I - \alpha  \cdot A} \right)^{ - 1}} - I,
\end{equation}
where $I$ is the identity matrix, $\alpha  > 0$ is the tunable parameter, ${\lambda _{max}}$ is the largest eigenvalue of $A$.
\par (5) Superposed Random Walk (SRW) considers the local random walk to emphasize the local nodes near starting point.
\begin{equation}\label{SRW}
  s_{xy}^{SRW}\left( t \right) = \mathop \sum \limits_{\tau  = 1}^t \left[ {{q_x}{\pi _{xy}}\left( \tau  \right) + {q_y}{\pi _{xy}}\left( \tau  \right)} \right],
\end{equation}
where ${q_x} = \frac{{{k_x}}}{{2\left| E \right|}}$  and ${\pi _{xy}}\left( \tau  \right)$ denote the transfer probability from $x$ to $y$.

\section{Results and discussions}
\begin{figure*}
\centering
  \subfigure[Hypertext]{
    \label{network1} 
    \includegraphics[width=3.1in]{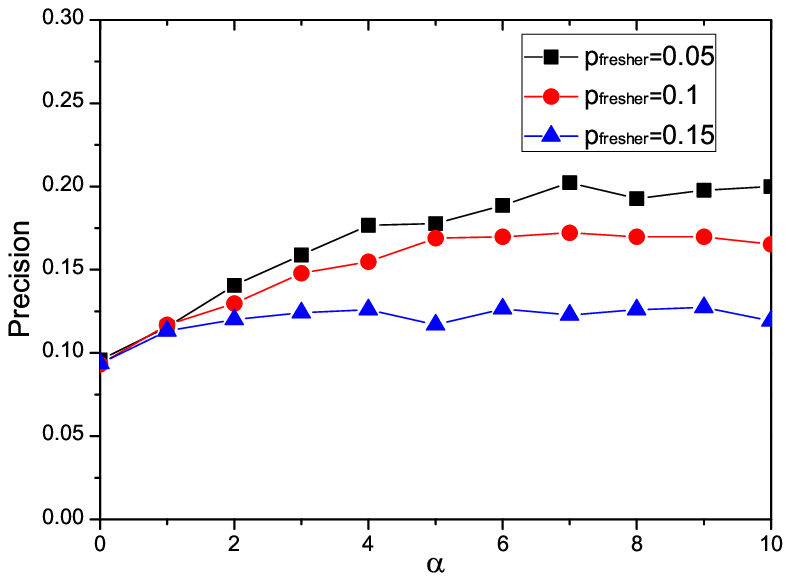}
    }
    \subfigure[Haggle]{
    \label{network1} 
    \includegraphics[width=3.1in]{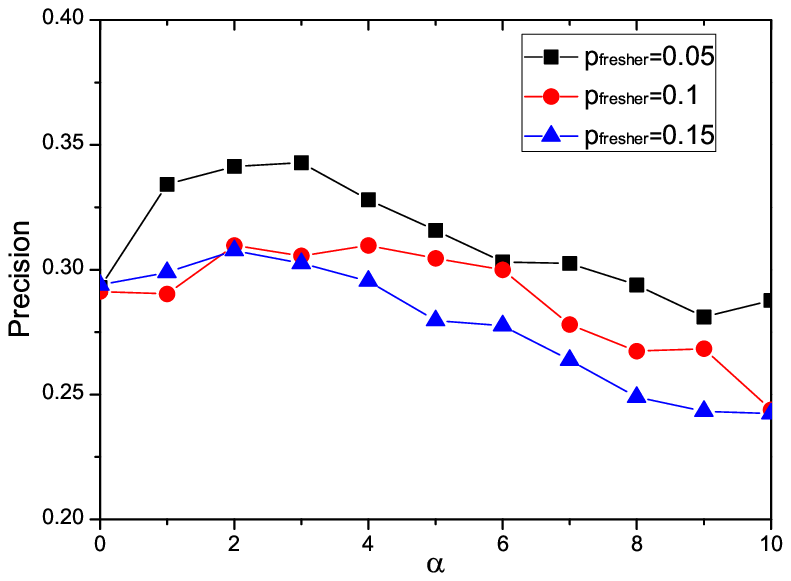}
    }
    \subfigure[Infec]{
    \label{network1} 
    \includegraphics[width=3.1in]{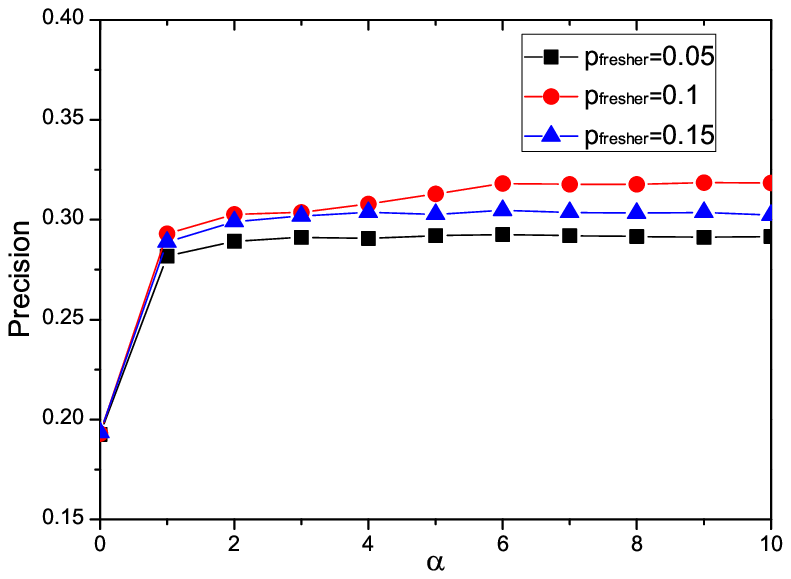}
    }
    \subfigure[UcSoci]{
    \label{network1} 
    \includegraphics[width=3.1in]{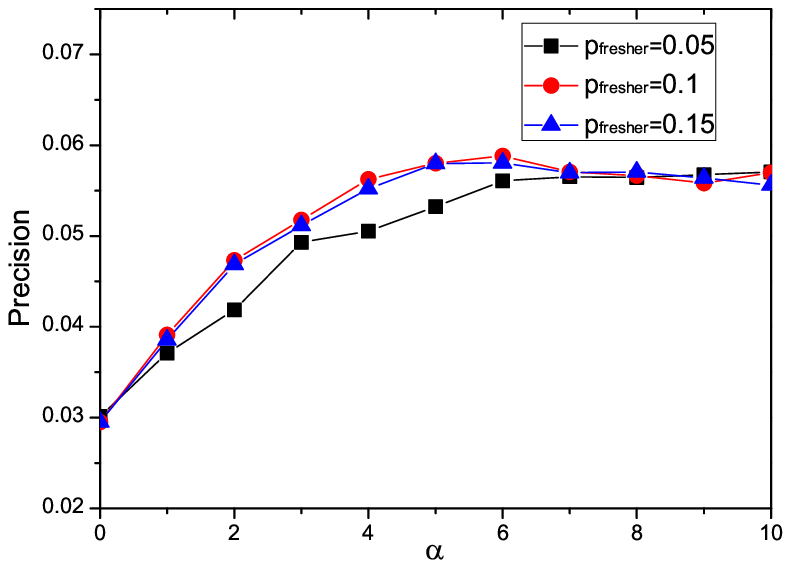}
    }
  \caption{(Color online) Precision versus $\alpha$ for $p_{fresher}=0.05, 0.10, 0.15$ based on PBSPM. The results here are averaged over 10 independent realizations in 4 temporal networks.}
  \label{precision}
\end{figure*}

 Popularity based structural perturbation method (PBSPM) integrates the attractiveness and popularity whose the effects on precision should be investigated firstly. With Eq .\ref{futureattractiveness}, PBSPM degenerates into original SPM when $\alpha=0$ since the popularity of all nodes are equal to 1. With the increase of $\alpha$, popularity starts to function in link prediction. And, given that the considered $p_{fresher}$ fraction of links has strong influence on final results, three precision curves of four real-world networks for typical values $p_{fresher}=0.05, 0.10, 0.15$ are plotted in Figure \ref{precision} to show the effects of popularity on precision for continuous range of $\alpha$, with $x$-axis labeled as $\alpha$ and $y$-axis labeled as Precision.
Benefiting from the popularity of nodes, all precision curves rise to different degree in contrast to SPM ($\alpha$=0). The value of $p_{fresher}$ and $\alpha$ that optimize the precision varies for different networks, 0.05 and 7 in Hypertext, 0.05 and 3 in Haggle, 0.10 and 6 in Infec, 0.10 and 6 in Ucsoci. In Figure \ref{precision}(a)(c)(d), precision curve rises rapidly when $\alpha$ is just combined, and then tends to keep stable after achieving the best. In Haggle, the false popularity of nodes are strengthened greatly with the exceeding $\alpha$ which weakens some attractive nodes incorrectly and results the lose of existent connections. Therefore, precision curve in Figure \ref{precision}(b) firstly increases but then decreases sharply. Moreover, even given the different $p_{fresher}$, the precision curves all show the similar trend and the curve for a certain $p_{fresher}$ in one network always presents the advantages in precision. For the small scale network Hypertext and Haggle, when $p_{fresher}=0.05$, most nodes are observed in fresh set to characterise the popularity. On the contrary, for Ucsoci and Infec, the size of fresher set is imbalanced in regard to the number of nodes when $p_{fresher}$ is small, thus, properly longer history information should be considered to describe the popularity better. Besides, in all networks, the precision is hampered over the optimal value duo to the enhancement of noise in fresh set, suggesting the influence of various $p_{fresher}$ on final results.

\begin{figure*}
\centering
  \subfigure[Predicted topology of SPM]{
    \label{network1} 
    \includegraphics[width=3.1in]{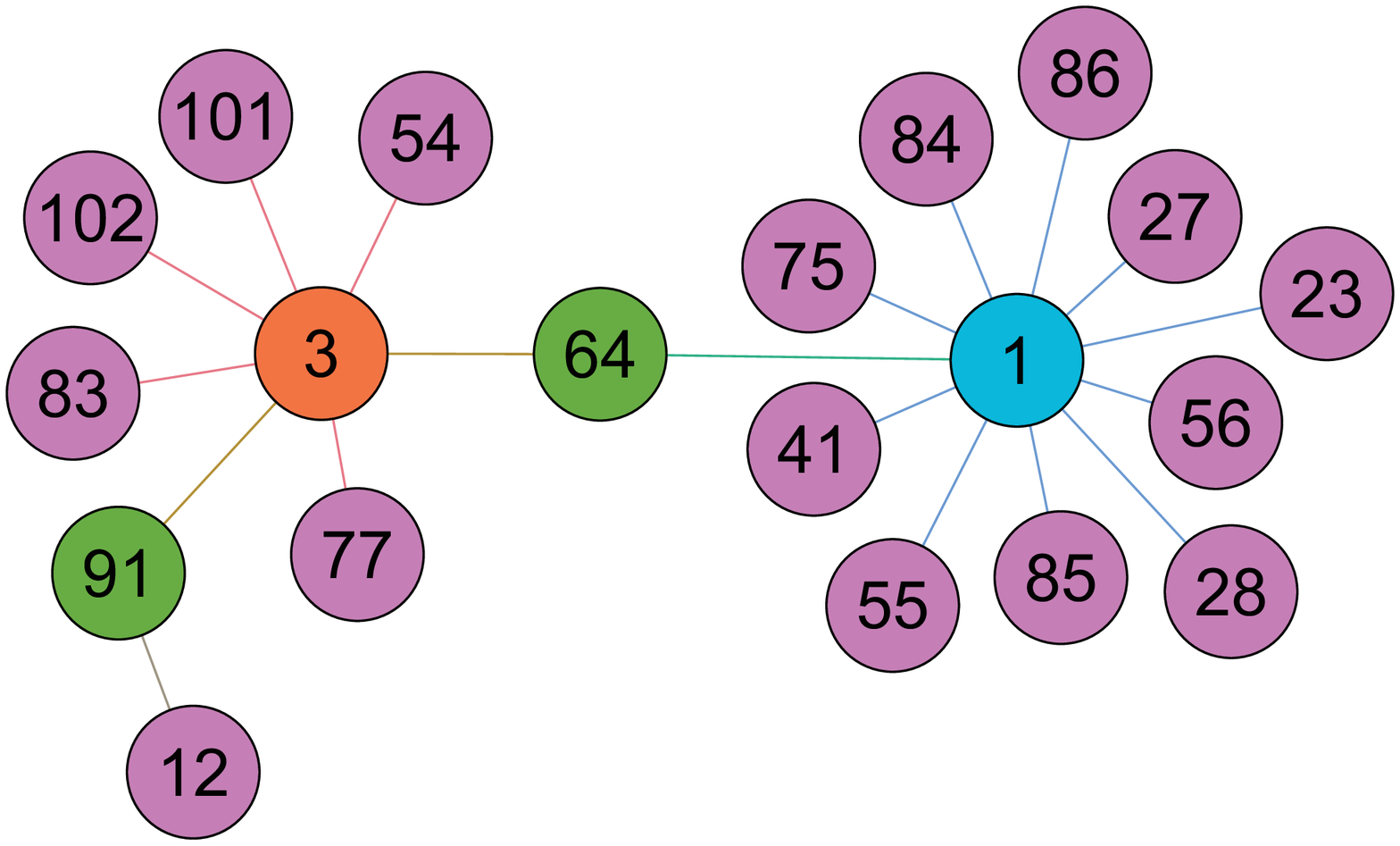}
    }
    \subfigure[Predicted topology of PBSPM]{
    \label{network1} 
    \includegraphics[width=3.1in]{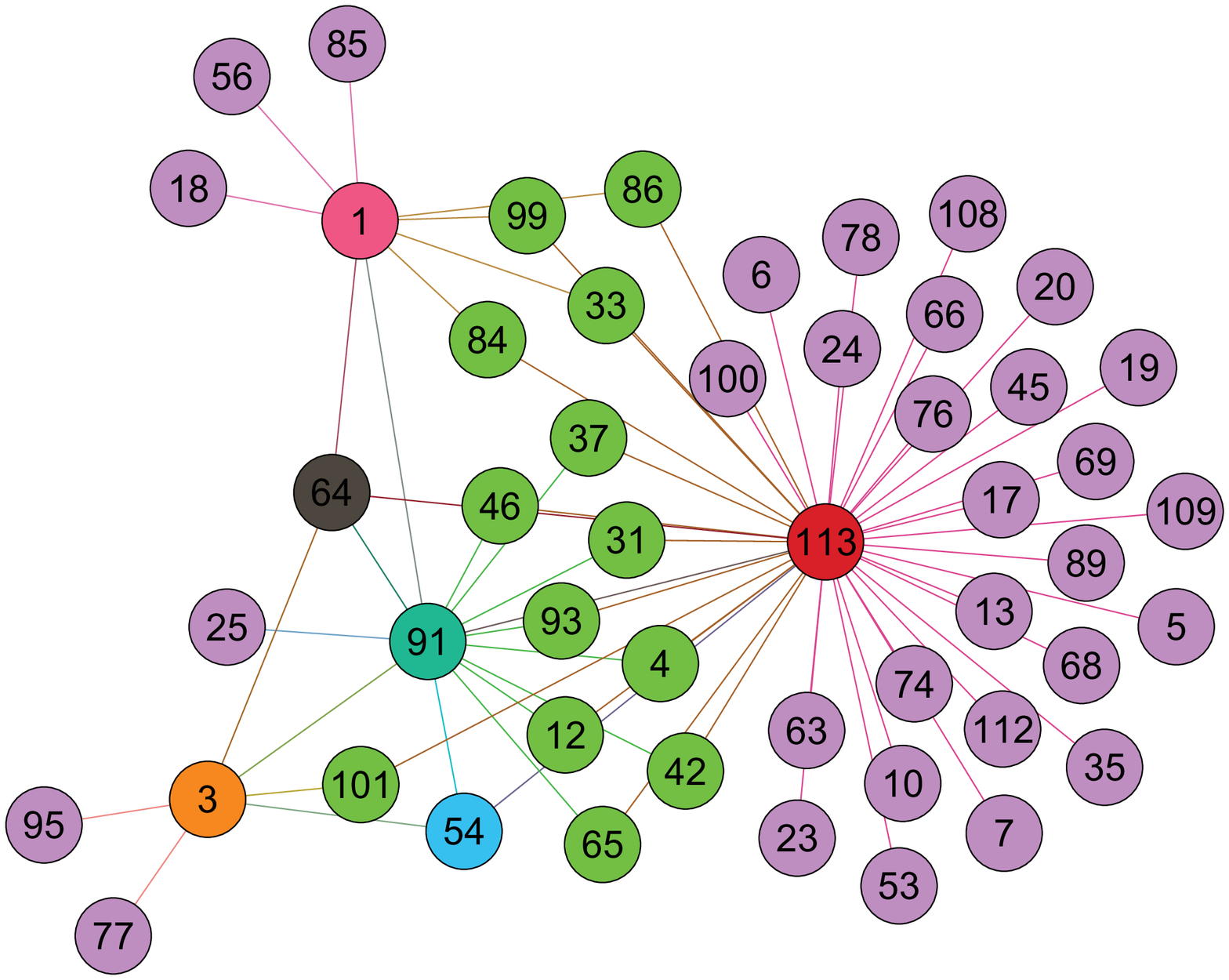}
    }
  \caption{(Color online) The edges of four typical nodes in Hypertext are predicted by SPM and PBSPM. The results of PBSPM are obtained with the parameters optimize the precision. }
  \label{topology}
\end{figure*}

\begin{figure*}
\centering
  \subfigure[Importance]{
    \label{network1} 
    \includegraphics[width=3.1in]{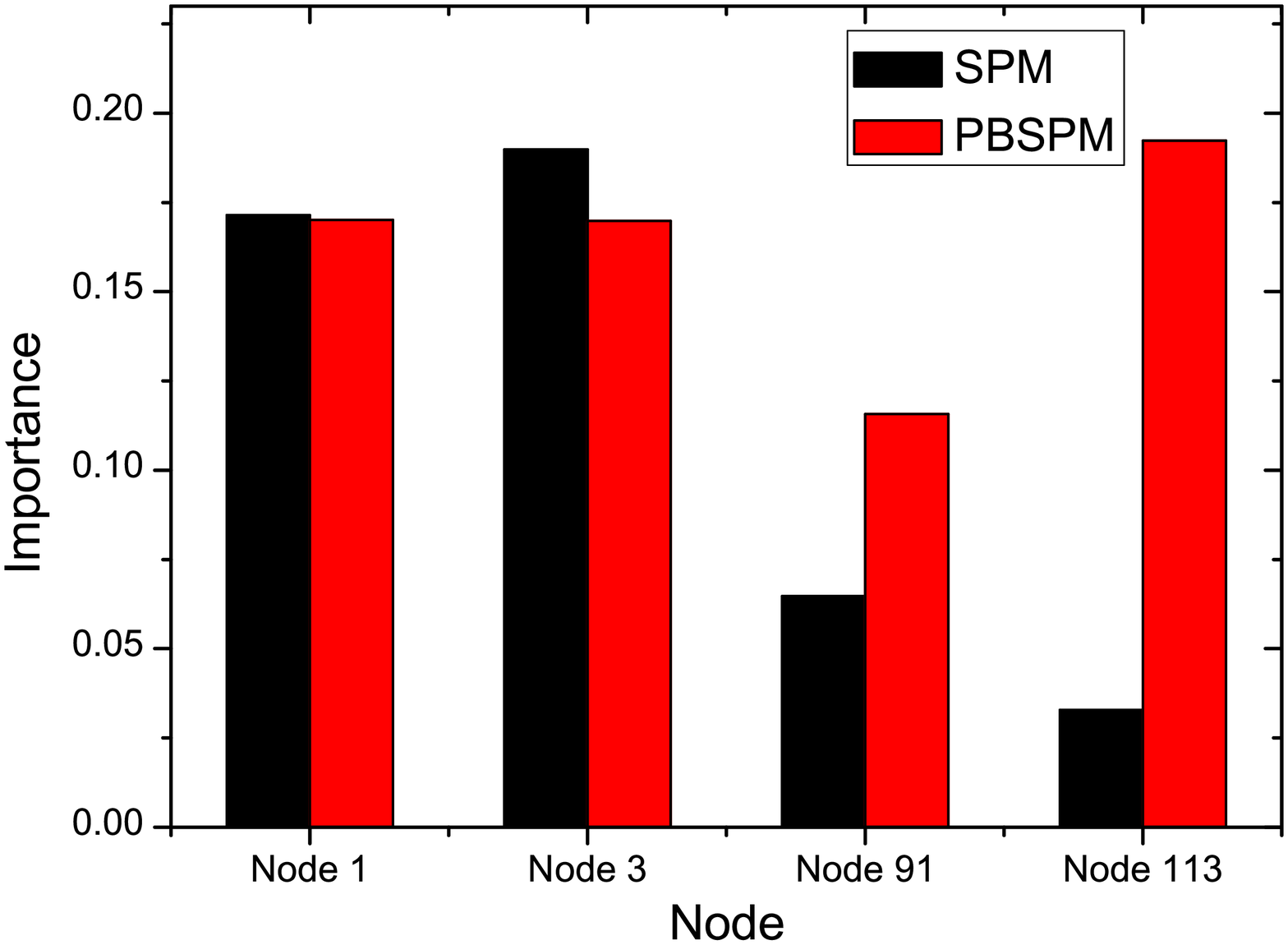}
    }
    \subfigure[Degree]{
    \label{network1} 
    \includegraphics[width=3.1in]{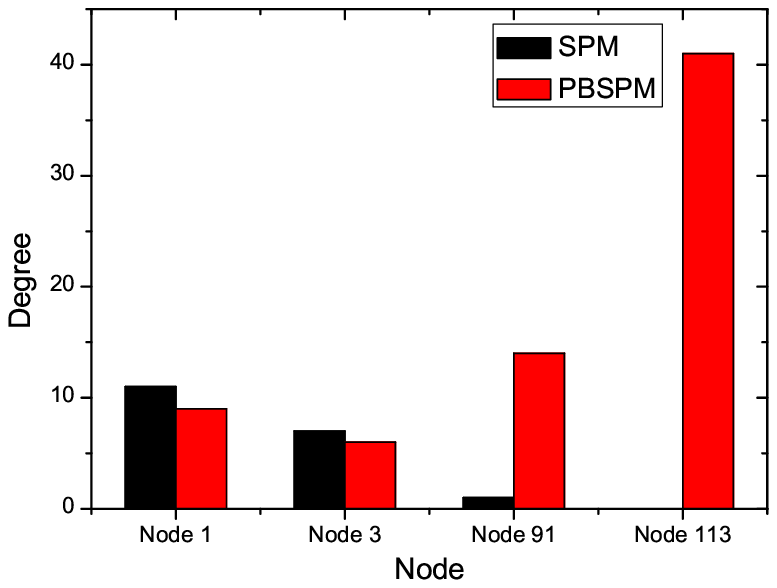}
    }
  \caption{(Color online) The original and advance importance of the four nodes are plotted as histograms corresponding to SPM and PBSPM. Degree of nodes are counted from Figure \ref{topology}.
  }
  \label{importance}
\end{figure*}

The validity of proposed algorithm in temporal networks has been confirmed. Urgently, the underlying reasons of these improvements are worth pursuing further. Taking example for Hypertext, after a random perturbation, the principal eigenvector $x_1$ for $A^R$ and the advanced $x'_1$ under the optimal case are calculated to quantify the attractiveness for the most weighted attractors. Notice that, the principal eigenvector not only characterises attractiveness of nodes, but also describes the importance. Four typical nodes are selected, the large-degree node 1 and 3, and the predicted active node 91 and 113, to analyse the predicted topology and corresponding variation of importance. Figure \ref{topology} plots the predicted topology of SPM and PBSPM respectively, and Figure \ref{importance} shows the importance and degree of nodes for above mentioned two cases with $x$-axis labeled as Node and $y$-axis labeled as importance and degree. In Figure \ref{topology}(a) without popularity, the large-degree nodes attract more links due to the higher importance, and few links connect to the unimportant nodes, especially, the connections of node 113 is 0 just for the lowest importance in Figure \ref{importance}(a). With popularity, the advanced importance results in the burst of links connecting to the two populary nodes in Figure \ref{topology}(b), particularly, degree of the most active node 113 ($s_{113}=1$) changes from 0 to 41. Reflected by Figure \ref{importance}, large degree nodes are always of higher importance and vice versa. As is expected, the importance of large-degree but inactive nodes are weakened to reduce the number of edges, otherwise, a node with higher popularity is enhanced via Eq .\ref{futureattractiveness} to attract much more links than before.

The above analysis focuses on the importance of several typical nodes. Definitely, the improvements result from the advanced attractiveness of all nodes. To comprehend the variation of attractiveness holistically, the Pearson correlation coefficient ($CC$) between principal eigenvector $x_1$ and degree increment in probe set are computed as follows:
\begin{equation}\label{cc}
  {cc = }\frac{1}{N}\sum\limits_{i = 1}^N {\left( {\frac{{{x_{1,i}} - \overline x_{1,i}} }{{{\delta _{{x_{1,i}}}}}}} \right)\left( {\frac{{{k_{i,probe}} - \overline k_{i,probe} }}{{{\delta _{{k_{i,probe}}}}}}} \right)},
\end{equation}
where $\overline x_{1,i}$ and $\overline k_{i,probe}$ are the means of and $s_i$ and $k_{i,probe}$. The $CC$ between advanced $x'_1$ and degree increment is obtained similarly. As is mentioned, principal eigenvector denotes the attractiveness for the most weighted attractor. Because $\left( {{{\lambda }_1} + {\Delta }{{\lambda }_1}} \right)\left( {x_1}{x_1}^{T} \right)$ occupies the main body of ${\tilde A}$. Hence, neglecting constant term ${{{\lambda }_1} + {\Delta }{{\lambda }_1}}$, similarity ${\tilde a}_{i,j}$ is mainly determined by eigenvector $x_1$. Table \ref{CC} lists the variation of Correlation Coefficient when popularity is considered and the averaged ${\Delta }{\lambda _1}$ after ten independent perturbations. $\Delta {CC}$ of four networks are all positive, suggesting attractiveness of some nodes are corrected to consist with the laws of nodes evolution. Crucially, the $\Delta {\lambda_1}$ is also positive, which strengthens the improvements of correlations. As a result, the active nodes are assigned more connecting opportunities to promote the precision.

\begin{table*}
\caption{\label{CC} $\Delta \lambda_1$ after perturbations and the variation of correlation coefficient $\Delta {CC}$ when popularity is considered, each result is averaged over ten perturbations. }
\begin{ruledtabular}
\begin{tabular}{p{1in}p{0.6in}p{0.6in}p{0.6in}p{0.6in}}
 Networks&Hypertext&Haggle&Infec&Ucsoci
\\ \hline
 $\Delta \lambda_1$&4.1822&4.79&1.86&4.2183\\
 $\Delta CC$&0.28&0.0582&0.3092&0.1155\\

\end{tabular}
\end{ruledtabular}
\end{table*}

\begin{figure*}
\centering
  \subfigure[Precision versus $m/n$]{
    \label{network1} 
    \includegraphics[width=3.1in]{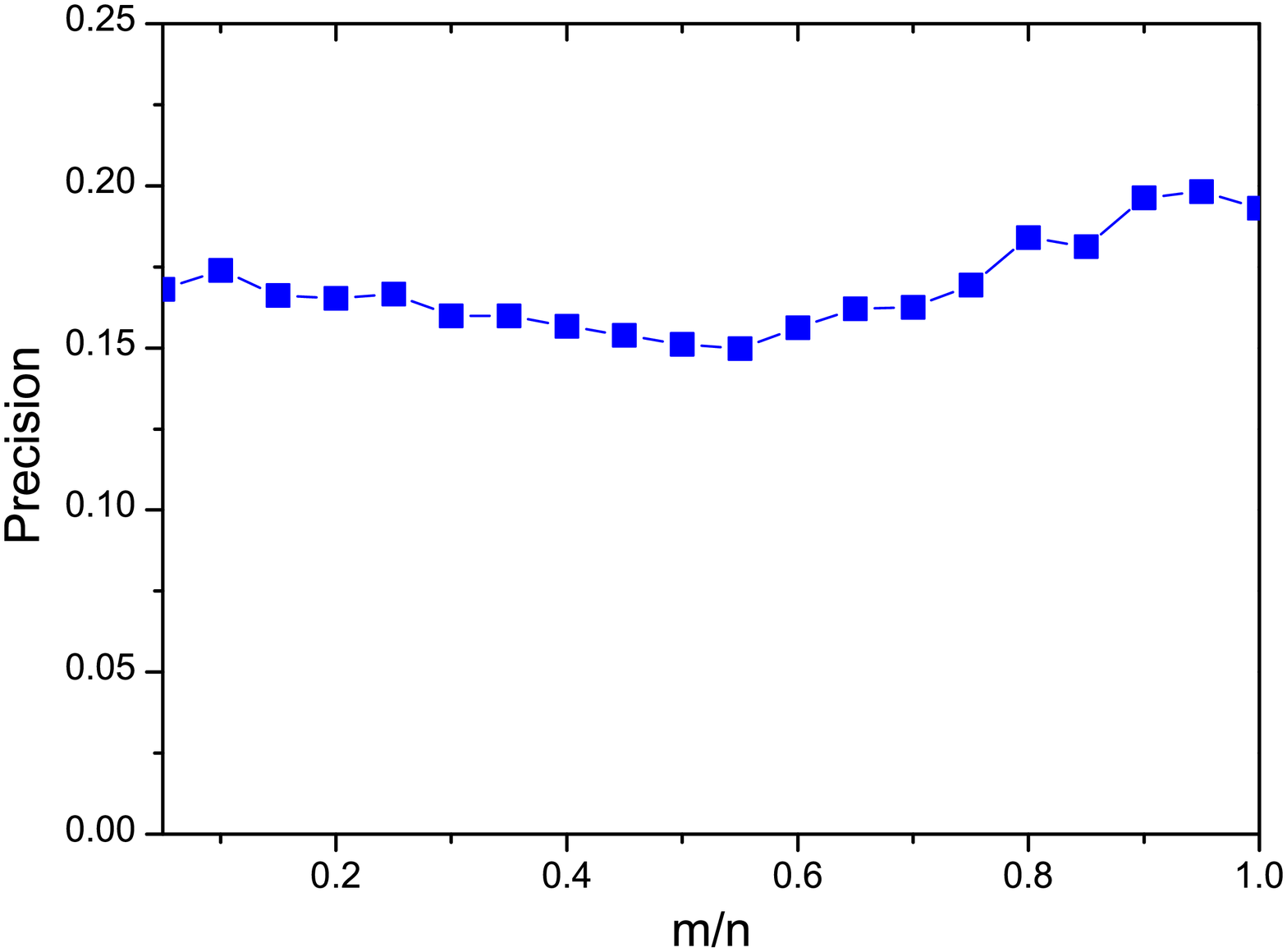}
    }
    \subfigure[The gap $\Delta\lambda_i$]{
    \label{network1} 
    \includegraphics[width=3.1in]{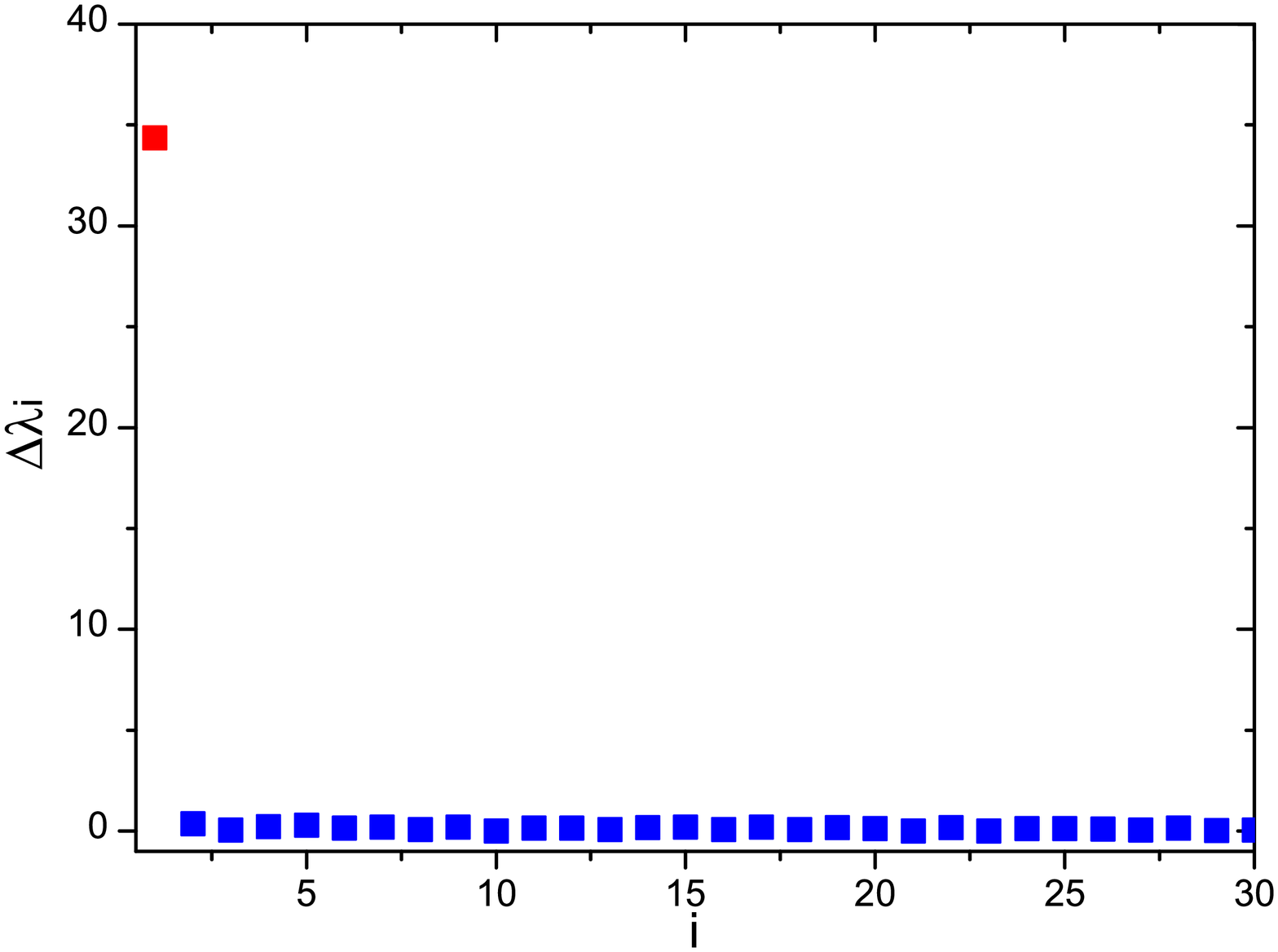}
    }
  \caption{(Color online) Precision versus $m/n$ and the gap $\Delta\lambda_i$ of $A^T$ for Hypertext. The precision curve is obtained with $p_{fresher}=0.05$ and $\alpha=7$. }
  \label{percentage}
\end{figure*}

Nevertheless, the high computation complexity of both SPM and PBSPM limits the application in large-scale networks. In real case, many common properties make no sense to characterise the similarity between two entities. Similarly, the attractor denoted by $x_k$ corresponding lower weight $\lambda_k$ could be treated as noise. Inspired by this, we propose a fast PBSPM that only considers a few more weighted attractors to reduce the computation complexity. Taking Hypertext as example, Figure \ref{percentage}(a) plots the precision for various $m$, with $x$-axis labeled as $m/n$ and $y$-axis labeled as Precision. Compared with SPM, the curve presents significant improvements with $0.05n$ most weighted attractors, then keep stable in a long interval, and ultimately achieves the best. The results meet the effectiveness of Eq. \ref{decompose2} and we suppose that there should be a $m<0.05n$ at which we can seek a well trade-off of computation complexity and accuracy. In practical networks, a huge gap usually exists between the large eigenvectors and the other eigenvalues of adjacent matrix $A^T$. Thus, eigenvectors corresponding large eigenvalues are emphasized while the others can be neglected roughly. To determine the optimal $m$ precisely, Figure \ref{percentage}(b) shows the gap $\Delta\lambda_i=|\lambda_i|-|\lambda_{i+1}|$ between the two eigenvalues, with $x$-axis labeled as $i$ and $y$-axis labeled as $\Delta\lambda_i$. The gap denoted by red point is distinct and the others including when $i>30$ are all close to 0, suggesting that the optimal value for Hypertext is $m=1$. For Haggle, Infec and Ucsoci, the $m$ are respectively determined as 2, 19, 2 after which the $\Delta\lambda_i$ approaches to 0 approximately. Later, Equation \ref{decompose2} with fixed $m$ is utilized and Table \ref{comparsion} shows the precision that indicates the fast PBSPM is a well trade-off of computation complexity and accuracy. Despite the slight lose of precision in some networks, the fast method still presents remarkable improvements compared with other predictors.

Lastly, Table \ref{comparsion} shows the optimal cases and precision of other traditional algorithms. Compared with SPM, both fast PBSPM and PBSPM achieve remarkable improvements in all networks, 112.09\% in Hypertext, 20.59\% in Haggle, 63.21\% in Infec, 97.32\% in UcSoci. The results in boldface suggest that the proposed methods outperform five state-of-the-art algorithms in accuracy. Besides, the two advanced SPM predict future links without the priori information of network organizations, improving the drawbacks of traditional methods and presenting high accuracy and robustness in evolving networks.

\begin{table*}
\caption{\label{comparsion}Comparison of accuracy measured by Precision of four networks. The results of fast PBSPM and PBSPM are obtained with the same $p_{fresher}$ and $\alpha$.}
\begin{ruledtabular}
\begin{tabular}{p{1in}p{0.5in}p{0.5in}p{0.5in}p{0.5in}p{0.5in}p{0.5in}p{0.5in}p{0.5in}}
 Precision &CN &AA &RA &Katz &SRW &SPM &Fast PBPSM &PBSPM
\\ \hline
 Hypertext&0.0959&0.1050&0.1005 &0.0959&0.1187&0.0984&\textbf{0.2087}&\textbf{0.2023}\\
 Haggle&0.1786&0.1888&0.1939&0.2041&0.2194&0.2928&\textbf{0.3531}&\textbf{0.3429}\\
 Infec&0.0233&0.1163&0.1814&0.0233&0.2884&0.1949&\textbf{0.3070}&\textbf{0.3181}\\
 UcSoci&0.0138&0.0153&0.0138&0.0138&0.0046&0.0298&\textbf{0.0587}&\textbf{0.0588}
\end{tabular}
\end{ruledtabular}
\end{table*}

\section{Conclusions}
In conclusion, the main contribution of our paper is to investigate the popularity of nodes in real-world evolving scenario and apply popularity to link prediction. Unlike previous work that calculate popularity with complex theory, we propose a simply approach to obtain popularity only based on straight statistics of training dataset. Then a hypothesis is proposed that current network structure is determined by history attractiveness of nodes, while the future network structure is further influenced by activeness of nodes. By introducing activeness into perturbation method, the proposed method could differentiate active and inactive history important nodes, and prefer to predicting new edges attaching active nodes with high history importance. Four real-world evolving networks is employed to test the performance of the proposed method. Comparing with traditional methods, PBSPM achieves better performance in precision. However the proposed popularity of a node has low correlation with future attracted edges, meaning that popularity metrics still need to improve. Improving popularity performance would enhance the precision of link prediction, which is the future task. Since our work mainly explore prediction in dynamical networks, it has extensive application in traffic prediction, airline control, recommendation of social network, and so on.

\section{Acknowledgement}
This work is jointly supported by the National Nature Science Foundation of China (Nos.
61004102, 61471243, 11547040 and U1301252), Science and Technolgy Innovation Commission of
Shenzhen (Nos. JCYJ20150625101524056, JCYJ20140418095735561, JCYJ20150731160834611 and
SGLH20131010163759789), China 863 (No. 2015AA015305) and Tencent Open Research Fund.



\end{document}